\documentclass{ws-rv975x65}
\usepackage{subfigure}
\usepackage{ws-rv-van}
\makeindex

\begin{document}

\chapter{ {\large \bf In Memoriam Nikolai Uraltsev} : \\    
Uraltsev's and other Sum Rules, Theory and Phenomenology of
$D^{**}$'s} 

\author{Alain Le Yaouanc, Olivier P\`ene}
\address{Laboratoire de Physique Th\'eorique, CNRS et Universit\'e Paris-sud XI,
b\^atiment 210, 91405 Orsay Cedex, France
\footnote{Unit\'e Mixte de Recherche 8627 du Centre Nationale de la Recherche
Scientifique.}.}
\vspace{0.3cm}


\begin{abstract}
We first discuss  Uraltsev's and other sum rules constraining the $B \to D^{**}(L=1)$ weak transitions in the infinite mass limit, and compare them with dynamical approaches in the same limit. 
After recalling these well established facts, we discuss how to apply infinite mass
 limit to the physical situation. We provide predictions concerning 
semi-leptonic decays and non-leptonic ones, based on quark models. We then present in more detail the dynamical approaches: the relativistic 
quark model \`a la Bakamjian-Thomas and lattice QCD. We summarise lattice QCD results in 
the infinite mass limit and compare them to the quark model predictions.
We then present preliminary lattice QCD results with finite $b$ and $c$ quark
masses. A systematic comparison between theory and experiment is performed. 
We show that some large discrepancies exist between different experiments. Altogether 
the predictions at infinite mass are in fair agreement with experiment
for non-leptonic decays contrary to what happens for semileptonic decays. 
We conclude by considering 
the prospects to clarify both the experimental situation, the theoretical 
one and the comparison between both.   
\end{abstract}

\section{In Memoriam of Kolya Uraltsev}
In november 2012 we had in Paris a workshop dedicated to $B \to D^{\ast\ast}$
decays. Many of the best specialists of the field were there, experimentalists
and theorists. We took time to discuss in detail about what we knew, what was
still obscure, how to solve the issues. Among the participants was Kolya Uraltsev.
He took a prominent part in the discussions. He was so deep, so rigorous, so
strong in his statements, and so careful about the yet unknown, about the
ambiguities, so positive about what should be done and about the requests to
experimentalists. 

We have been working for long on the heavy quark limit for the $B \to
D^{\ast\ast}$ decay using a relativistic quark model.  In 2000 Kolya derived a
very simple and powerful sum rule in the same limit. From there on we started to interact with
him. Ikaros Bigi came several times in Orsay and we have published together two
``memorinos". Bigi was a close collaborator and friend of Kolya. Thanks to him
we could communicate more intensively with Kolya. All these discussions were
really enlightening. 

Kolya did not come to our workshop dinner arguing that he would be tempted by
the food and that it was not good for him. We did not realise the case was so
serious. In February 2013 we received a mail from Ikaros ``Dear Friends, 
on Wednesday (February 13th 2013) Kolya had passed away in Siegen -- 
he was a wonderful person, theorist and true friend. 
I cannot speak more.   Ikaros". We couldn't say more either, the shock was 
too violent. 

Now the time has come to honour Kolya by speaking about physics, which was so
important for him, to which he devoted all his strength and his admirable brain.   
\section{ Historical elements concerning the discussions about $D^{\ast\ast}$} 

The issue of the decays $B \to D^{\ast\ast}$ has been actively discussed since
more than twenty years under its theoretical and experimental aspects and the
relation between both, see~\cite{memorino1,memorino2}  for an overlook on these
aspects and more recently~\cite{proposal}.

\subsection{Well established facts at $m_Q \to \infty$} \label{well}

Let us first recall what are the $D^{**}$ under concern. They are the charmed
states with quark model assignment $L=1$, and most often we are meaning the
lowest lying ones. In the infinite mass limit $m_Q \to \infty$, they can be
separated into two doublets $j=1/2,j=3/2$, then four states, almost degenerate
(except for a small spin orbit force), for each level of excitation. While the $j=1/2$ are broad, the $j=3/2$ are narrow, having
respectively $S$ and $D$ waves in pionic decay.  The broad $j=1/2$ include a
$0^+$ and a $1^+$, while $j=3/2$ include a $2^+$ and a $1^+$.

These states are playing a prominent role in the heavy quark sum rules, e.g. the
Bjorken sum rule:
\begin{eqnarray}
\rho^2=1/4+\sum_n (|\tau_{1/2}^{(n)}(w=1)|^2+2~|\tau_{3/2}^{(n)}(w=1)|^2)
\end{eqnarray}
$n$ labels successive levels of excitation. $\rho^2$ is the forward slope of the
elastic Isgur-Wise function, while  $\tau_{1/2,3/2}$ are similar quantities for
the $B \to D^{**}$ transitions~\cite{Isgur:1990jf} . One defines for 
instance for the $B \to D(0^+)$ transition:
\begin{eqnarray}
\langle 0^+~|A_{\lambda}|~0^- \rangle=-
\frac {1}{\sqrt{v_0v_0'}}(v_{\lambda}-v_{\lambda} ')~\tau_{1/2}(w)\label{tau}
\end{eqnarray}
where $v,v'$ are the four-velocity~\footnote{The four-velocity is defined 
by $v_\mu \equiv p_\mu/m$.} vectors of the mesons. $w=v.v'$, $w=1$ corresponding
to zero recoil. We had observed that this rule is exactly satisfied in a class
of quark models \cite{duality}. 

In 2000, N. Uraltsev \cite{Uraltsev:2000ce} discovered a new class of sum rules,
similar but for the combination of $\tau$'s, which appear now in differences,
e.g.:
\begin{eqnarray}
\sum_n (|\tau_{3/2}^{(n)}(w=1)|^2-|\tau_{1/2}^{(n)}(w=1)|^2)=1/4 \label{ur}
\end{eqnarray}
We had found, also from duality arguments,  one similar 
rule~\cite{LeYaouanc:2001vu}, except for factors of excitation energy, i.e.  the
sum has now factors of energy minus the ground state energy :
\begin{eqnarray}
\sum_n (E_{3/2}^{(n)}-E_0)|\tau_{3/2}^{(n)}(w=1)|^2-(E_{3/2}^{(n)}-E_0)|\tau_{1/2}^{(n)}(w=1)|^2)=...
\end{eqnarray}
It is retrieved by the systematic approach of N.Uraltsev in a whole series of
sum rules with increasing powers of these energy differences.

At this point, one enters into what will be the main worry in the rest of the
chapter. Both rules were indicating an important trend for the $\tau$'s : that
one should have $|\tau_{3/2}|$ on the whole larger than $|\tau_{1/2}|$. This is
yet a vague statement, which receives a stricter form under the assumption,
put forward by N.Uraltsev, that the lowest level (noted $n=0$) is dominating the sums. Then :
\begin{eqnarray}
|\tau_{3/2}^{(0)}(w=1)|>|\tau_{1/2}^{(0)}(w=1)|\label{1sur3}.
\end{eqnarray}
One must be aware that for the deduction to be valid, the dominance must be
rather strict.

The statement \eref{1sur3} is precisely the conclusion we had obtained , in the heavy quark limit, from our
Bakamjian-Thomas (BT) relativistic quark model approach in the heavy quark limit
(shortly described in \sref{secBT}), some years before. 

Another, related, point is that, combining the Uraltsev sum rule~\eref{ur}, with
the Bjorken one, one gets a stronger bound on $\rho^2$:
\begin{eqnarray}
\rho^2 >3/4
\end{eqnarray}
We had also observed previously that precisely this bound is obtained in the
``Bakamjian-Thomas" approach. And, in fact, we have also shown that the Uraltsev sum
rule is exactly satisfied in the model \cite{uraltsev-nous}.

One sees easily that the difference between $\tau_{3/2}(1)$ and $\tau_{1/2}(1)$
is a relativistic effect. Indeed, in an expansion in terms of the light quark
internal velocity, $v/c$, $\tau_{3/2}(1)$ and $\tau_{1/2}(1)$ are of order
$c/v$, they have the form of dipole transitions where the current matrix element
is $\propto \vec{r} \propto m~c/v$ and does not depend on the spin. Then, in
the non-relativistic limit $\tau_{3/2}(1)-\tau_{1/2}(1) = 0$. The difference is
subleading in the expansion, and the Uraltsev sum rule displays this : $1/4$ is
subleading. But why is the difference positive ?

This type of quark model offers a very simple and intuitive explanation of the
difference: one ends with the simple expression
\begin{eqnarray}
\tau_{3/2}(1)-\tau_{1/2}(1)={1\over 2}\int_0^\infty dp \ p^2\ \phi_1(p)
 \frac {p} {p_0+m} \phi_0(p)
\end{eqnarray}

\par \vskip 3 truemm
  $\phi_{L}(p)$ are the radial wave functions. For simplicity, we have taken
 $\phi_{1}(p)$ the same for all four states,
which amounts to neglect a small spin-orbit force. The two functions
$\phi_0(p),\phi_{1}(p)$ have been chosen positive by a choice of the phase of
states, and the $\tau$'s are then positive. This choice is possible for the
lowest states, but not for the $L=1$ radially excited states, for which there
are zeroes in the radial wave functions.
\par \vskip 3 truemm
The ${\cal O}(v/c)$ factor  $\frac {p} {p_0+m}$ comes from the Wigner rotation
 of the light (spectator) quark spin which acts differently on $(1/2)^+$ and
 $(3/2)^+$  states. This Wigner rotation is  a typical relativistic effect in
 bound states~\cite{Morenas:1997nk}.
\par \vskip 3 truemm
It is of order ${\cal O}(v/c)$, which gives indeed ${\cal O}(v/c)^0$ for the
difference of squares in Uraltsev sum rule (recall that each term squared is
${\cal O}(v/c)^{-2}$). This is compatible with the $1/4$ in the r.h.s of 
\eref{ur} : $\tau_{3/2}(1)-\tau_{1/2}(1)$ is {\bf positive}. 
Moreover it is large in fact, because the velocity of {\bf light} quarks within
 mesons is large : relativistic internal quark velocities . 
 
In summary, $\tau_{3/2}(1)-\tau_{1/2}(1)$ is {positive} and {large}. Or, as
found quantitatively, $\tau_{1/2}(1)$ is small with respect to $\tau_{3/2}(1)$.
In fact, using what is the best potential in our opinion, the model of Godfrey
and Isgur~\cite{Godfrey:1985xj}, in the ``Bakamjian-Thomas" formalism, we found : 
\begin{eqnarray}\label{eq:BTres}
\tau_{3/2}(1)&=& 0.55 \\
\tau_{1/2}(1)&=& 0.225 
\end{eqnarray}
Of course, the quark model is not meant to be an exact approach.\footnote{We do
not quote errors however.  This simply means that we cannot calculate errors.
Even quoting a range of variation when one varies some parameters of the model
could be misleading; one would have to check that the new values of the
parameters fit the spectrum and the other reference data as well, which is not
done usually. What can be said usefully is that the $\tau_{1/2}$ is more
sensitive to the details of the potential model at short distance.}.

Only a true QCD calculation can allow a safe conclusion, therefore  it is very
important that the conclusion was indeed confirmed later in lattice QCD, at
$N_F=0$ and then at $N_F=2$. The latest result is in \eref{final}.

One notes that the agreement of the quark model with lattice is very good for
$j=3/2$ (central lattice value $\simeq  0.53$), but that it is slightly worse for $j=1/2$~: with the lattice central
value ($\simeq 0.3$), quark model is sizeably below, and in squares, which are relevant in the
above sum rules and for the rates, the ratio between $\tau_{1/2}(1)$ from lattice
and the same from the relativistic quark model is not far from $2$ : $1.7$.
Still the close agreement between these estimates for $\tau_{3/2}(1)$ and the
semiquantitative agreement for $\tau_{1/2}(1)$ is striking and gives us 
confidence in these figures. 
   
The advantage of lattice QCD is also that one can estimate errors, since one has
now a systematic approach to the true result of QCD. By successive improvement
of the calculation, one can estimate the errors with respect, for example, to
the continuum and chiral limits.  

Up to now everything seems consistent and very encouraging. Lattice approach,
general heavy quark statements and quark model agree. The lacking protagonist
is experiment. And now comes a possible trouble.

\subsection{Phenomenology in the $m_Q \to \infty$ approximation }\label{historical}

The results in the infinite mass limit can be used in two ways :
\begin{itemlist}
\item either as purely theoretical tests of the consistency of various
approaches-e.g. does one get in a model the necessary Isgur-Wise scaling and
the normalisation condition $\xi(1)=1$ ? Or, for the lattice approach, where it
is not the question, it may be still a practical check of the soundness of the
calculation.
\item or, more ambitiously, it could be relevant for phenomenology, as well as model
dependent results like the ones from the quark model, or exact dynamical
calculations of lattice QCD, taken in the the $m_Q \to \infty$.   
\end{itemlist}
Here, the initial idea is of course that for heavy flavor physics the heavy
quarks $c,b$ could be sufficiently heavy for the heavy quark limit to be
applied, and that many advantages are obtained in that limit. First, new general
statements and simplifying features, then new properties of the quark model and
other approaches. Even for lattice QCD, the $m_Q \to \infty$ limit is useful
practically because the treatment of the heavy quark line is especially easy
(the heavy quark being described by a Wilson line instead of having to solve
numerically a Dirac equation).

A striking example of the phenomenological success of the approach is the
prediction of broad and narrow $L=1$ $D^{**}$ states, the $j=1/2$ having pure
$S$ wave pionic decay, while the $j=3/2$ have it in $D$ wave. One also observes
an agreement for $B \to D^{(*)}$ and for $B \to D^{**}(j=3/2)$ (narrow
states). In the last case, this is especially true if one averages over the $j$
multiplet.

Then, one would expect also the above hierarchy $j=1/2 \ll j=3/2$ to be 
observed. In fact, there is at present no clear conclusion, after twenty years
of experimental effort.

At this stage however, it must be said more precisely how the $m_Q=\infty$
predictions, which are well defined by themselves, may be used for real physics,
at finite masses. 

There is admittedly no compelling procedure. The only real safe way would be to
calculate at the finite physical masses, but we miss theoretical tools to do so,
or to calculate corrections to the infinite mass limit. Precisely, the hope was
in the beginning to simplify the problems by a clever use of the
$m_Q=\infty$ limit \footnote{In addition, at finite mass, one enters new
difficulties for the quark model, like loss of covariance or non conservation of
the vector current.}. 

Then, one is just proposing recipes, which have succeeded in the above cases $B
\to D^{(*)},D^{**}(j=3/2) $. To be specific, among the many choices of form
factors, what are the ones that we assimilate to their $m_Q=\infty$ limit ?
Different choices will lead to different results for the physical predictions
from the same heavy quark theory.

In our papers, we have made a choice based on \eref{tau}. This means that
one keeps the same expression, but with physical velocities. There is at least
some logic in using invariant form factors defined with velocity coefficients.
The exact formula at finite masses duely contains two independent form factors,
which one defines also with velocity factors :
\begin{eqnarray}
\langle 0^+~|A_{\lambda}|~0^- \rangle=\frac {1}{2\sqrt{v_0v_0'}}((v_{\lambda}+v_{\lambda} ')~g_+(w)+
(v_{\lambda}-v_{\lambda} ')~g_-(w)) \label{ff0plus}
\end{eqnarray}
with $g_+(w) \propto 1/m_Q$ and $g_-(w) =-2~\tau_{1/2}(w)+ {\cal O}(1/m_Q)$.
What we do is to cancel plainly the ${\cal O}(1/m_Q)$ in $g_{+,-}(w)$, i.e. we
set $g_{+}(w)=0,g_{-}(w)=-2~\tau_{1/2}(w)$.

It is useful to quote the differential rate corresponding to \eref{ff0plus}  
for later discussion:
\begin{eqnarray}
\frac {d\Gamma} {dw}=|V_{cb}|^2 \frac {G_F^2~m_B^5}{48 \pi^3} r^3(w^2-1)^{\frac {3}{2}} [(1+r)g_+-(1-r)g_-]^2 \label{rate0plus}
\end{eqnarray}
with $r=m_D^{**}/m_B$.
\subsection{Semileptonic decays ; the beginning
of controversies about the $j=1/2$ states} \label{SL}

The predictions have been formulated for integrated rates. Indeed, the 
semileptonic data were not initially sufficiently accurate to make a comparison for
differential distributions.  Note that, for the safer lattice QCD,  at
$m_Q=\infty$, one is restricted to $w=1$ and the semileptonic data are not
sufficiently accurate to make a comparison in differential distributions around
$w=1$ even now. Therefore, one is led to use the quark model even presently.

One obtains for the semileptonic rate of $j=3/2$ in the above mentioned
quark model~\cite{Morenas:1997nk}, with the choice of Godfrey and Isgur as the best potential model :
\begin{eqnarray}
BR(B \to D^{(\ast\ast)}(j=3/2)2^+\,l\,\nu)&=& 0.70 \, \%\qquad D^\ast_2 \\
BR(B \to D^{(\ast\ast)}(j=3/2)1^+\,l\,\nu)&=& 0.45 \,\% \qquad D_1 \\
BR(B \to D^{(\ast\ast)}(j=3/2)\,l\,\nu)&=& 1.15 \, \%\label{sum32}
\end{eqnarray}
while for $j=1/2$:
\begin{eqnarray}
BR(B \to D^{(\ast\ast)}(j=1/2)1^+\,l\,\nu)&=& 0.07\,\%\qquad D^\ast_1 \label{1+} \\
BR(B \to D^{(\ast\ast)}(j=1/2)0^+\,l\,\nu)&=& 0.06\,\% \qquad D^\ast_0\\ 
BR(B \to D^{(\ast\ast)}(j=1/2)\,l\,\nu)&=& 0.13\,\%
\end{eqnarray}
We have indicated on the right the identification of the considered 
$D^{(**)}$ with the experimental states in the notations of \cite{ref:pdg}, terms
 although the notations with $j=1/2,3/2$ seem to us much more transparent.
Note that the inequality $BR(j=1/2) \ll BR(j=3/2)$ is much stronger that the one
between $\tau$'s, first because one has squares, second because there is an
additional kinematical factor in the rates.
\begin{eqnarray}
\frac{d\Gamma_{1/2}/dw} {d\Gamma_{3/2}/dw}= \frac {2} {(w+1)^2} \left(\frac {\tau_{1/2}(w)}{\tau_{3/2}(w)}\right)^2 
\end{eqnarray}
where one has summed within the $j$ multiplets. The kinematical ratio $\frac {2}
{(w+1)^2}$ adds a factor $< 1/2$ to the ratio $\left(\frac
{\tau_{1/2}(w)}{\tau_{3/2}(w)}\right)^2$.

While $j=3/2$ states are in rough agreement with experiment from the beginning,
especially the sum, $j=1/2$ ones were strongly disagreeing  with the results of
Delphi, table 10 in~\cite{delphi} at LEP : a large signal was found there for a broad $1^+$,
an order of magnitude larger than the prediction in \eref{1+} : $BR(B \to
D^\ast_1 \,l \,\nu)=(1.25\pm 0.025 \pm 0.027) \%$. The $0^+$ was not clearly
seen : $BR(B \to D^\ast_0\, l \,\nu)=(0.42 \pm 0.33 \pm 0.22) \%$\footnote{This
number differs from the one quoted in the memorinos \cite{memorino1,memorino2}.}.

We have noted that the lattice QCD number for $\tau_{1/2}(1)$ is somewhat higher than
the quark model. Therefore, one would expect larger predictions for $BR(B \to
D^\ast_1 \,l \,\nu)$ by a factor around two, which would still leave the Delphi 
data unexplained.

Of course, the discrepancy could originate in the $m_Q \to \infty$ approximation.
Therefore, very early, general estimates of ${\cal O}(1/m_Q)$ corrections have
been given by Leibovich et al.~\cite{Leibovich:1997em}, in a very careful 
discussion. Concerning $j=1/2$, as well as the $j=3/2, 1^+$ they show that the vanishing of the
amplitude  at zero recoil which happens in the infinite mass limit is no longer
valid at finite masses, while for the $j=3/2, 2^+$, vanishing at zero recoil stays
valid at finite masses (it has a general kinematical origin). Then they predict very small enhancement for the $j=3/2, 2^+$. The enhancement they predict by an
estimate of the subleading $1/m_Q$ contributions does not exceed a factor 3 for
 $j=3/2,1^+$ and $j=1/2,0^+$ and is numerically very small for $j=1/2,1^+$. 
Their estimate is of course somewhat model dependent. 

For the $0^+$, one can understand rather simply that a large enhancement is possible from 
\eref{rate0plus}, through a series of effects ( following the discussion in
\cite{finite}). Now $g_+ \neq 0$. The contributions from $g_-$ and $g_+$ add
algebraically and may be interfering constructively as found from HQET for first
order correction at zero recoil and also in the quark model. The contribution
of $g_-$ is maintained close to its $m_Q \to \infty$  limit according to the
quark model, while $g_+$ becomes sizable (in fact the BT model finds it rather
large, but it is doubtful since it outpasses the ${\cal O}(1/m_Q)$ magnitude
 predicted by HQET near
zero recoil). The $g_+$ contribution is enhanced relatively to the latter by the
factor $(1+r)/(1-r)$. All these effects are squared in the rate. The predicted 
enhancements deserve consideration, but do
not correspond to the conclusions of DELPHI, which point to  a very large $1^+,1/2$.  
We will return to this issue of finite mass corrections when
speaking about lattice results at finite masses. 

Unexpectedly, the new experiments at the two B factories have not clarified the
experimental situation. Quite on the contrary : they have added new
contradictions~: contradictions within experiment for the same semileptonic
reaction, and others between the semileptonic and the non leptonic transitions,
newly measured (see below). And the latter were found rather in agreement with
the expectations from the $m_Q \to \infty$ approximation.  
\subsection{Nonleptonic decay. Class I and class III} \label{phenNL}

Indeed, a new, very interesting piece of data has appeared with the measurement of
the $B \to D^{**}$ transitions, first by CLEO. Namely the quasi-two body
decays $B \to D^{**} \pi$.  And it has been realised first from a paper by
Neubert \cite{neubert} that there was a close connection with the semileptonic
decay. Indeed through ``factorisation" (i.e. the old SVZ ``vacuum insertion"), the
$B \to D^{**} \pi$ amplitude can be related to the $B \to D^{**}$ current matrix
element. 
\begin{equation}
B \to D^{**} \pi= f_{\pi} p_{\pi}^{\mu} <D^{**}|J_{\mu}|B>
\end{equation}
and then the latter to the semileptonic
amplitude. In fact the $B \to D^{**} \pi$ rate is found to be :
\begin{equation}
   \frac{\Gamma(\bar B^0\to H_c^+\pi^-)}
   {{\mathrm d}\Gamma(\bar B^0\to H_c^+\ell^-\bar\nu)/
    {\mathrm d}q^2 \Big|_{q^2=0}}
   = 6\pi^2 f_\pi^2\,|V_{ud}|^2\,|a_1|^2
   + O\left( \frac{M_\pi^2}{M_B^2} \right)
\label{fact}
\end{equation}
where $H_c$ is any charmed meson, and where $a_1$ is very close to 1.
The statement is approximate because factorisation is only an (intuitive and
uncontrolled) approximation. Moreover, the formula~\eref{fact} applies properly
to the decay $B^0 \to D^{**-} \pi^+$. For $B^+ \to \bar D^{**0} \pi^+$, there is
another important diagram contributing, obtained by Fierz transformation, with
$B \to \pi$ current matrix element and the emission of the $\bar D^{**0}$
described by the annihilation constant $f_{D^{**}}$. One distinguishes the two
type of decays as being repectively ``class I" and ``class III" decays. Only class
I decay is directly calculable from the  $B \to D^{**}$ current matrix element.
These decays are discussed in detail in~\cite{jugeau}.

Let us then first consider this simplest case of class I. The predictions of the
quark model with Godfrey-Isgur wave functions in the $m_{Q} \to \infty$
approximation are then~\cite{proposal}~:
\begin{eqnarray}
\label{NL1}
&&B\left ( \overline{B}^0 \to D_2^{3/2\ +}\pi^-\right ) = 11 \times
10^{-4} \nonumber \\
&&B\left ( \overline{B}^0 \to D_1^{3/2\ +}\pi^-\right ) = 13 \times 10^{-4} \nonumber
 \\
&&B\left ( \overline{B}^0 \to D_1^{1/2\ +}\pi^-\right )= 1.1\times 10^{-4}\\
&&B\left ( \overline{B}^0 \to D_0^{1/2\ +}\pi^-\right )=1.3 \times 10^{-4} \nonumber \label{GIclassI} . 
\end{eqnarray}

This time, in contrast with the semileptonic case, they were found in rather
good agreement with the Belle experiment, the first to give an analysis of
the four states. In particular, the striking dissymetry $BR(j=1/2) \ll
BR(j=3/2)$ was observed. Finite mass corrections have been estimated according 
to the procedure of Leibovich et al. and one has found conclusions similar to
the semileptonic case~\cite{jugeau}. Namely, only the $0^+$ is expected to be
 affected by a large possible enhancement. Rather recently, the $0^+$ was also 
measured by Babar, with a larger rate but much larger errors, then not modifying 
the initial conclusion, see~\tref{tab:dsstarpi}

Then about class III. There comes one more statement in the $m_{Q} \to \infty$
limit : $f_D^{**}(j=3/2)=0$ while $f_D^{**}(j=1/2)$ is not
all suppressed. A rough estimate shows that for $j=1/2$, the contribution with
$f_D^{**}$ is large and wins over the class I, which is suppressed, and
that it has the same sign~\cite{jugeau}. 

A consequence of this is that class III should be almost the same as class I for
$3/2$, but differ strongly for $1/2$. The suppression of $1/2$ predicted and
observed in class I should not be present in  class III. An estimate from the
same paper~\cite{jugeau} is :
\begin{eqnarray}
\label{NL2}
&&B \left ( B^- \to D_2^{3/2\ 0} \pi^-\right ) = (8.7 \pm 3.2) \times 10^{-4} \nonumber \\
&&B \left ( B^- \to D_1^{3/2\ 0} \pi^-\right ) = (10.2 \pm 2.3) \times 10^{-4}\nonumber \\
 &&B \left ( B^- \to D_1^{1/2\ 0} \pi^-\right ) = (7.5 \pm 1.7) \times 10^{-4}\nonumber\\
 &&B \left ( B^- \to D_0^{1/2\ 0} \pi^-\right ) = (9.1 \pm 2.9)\times 10^{-4}\ .  \end{eqnarray}
This agrees with what is observed~\tref{tab:dsstarpi}.

It is now worth describing and discussing in more details the two dynamical
approaches at our disposal, with their properties at physical masses as well as
at $m_Q=\infty$, and then the present experimental situation.
\section{Quark model}

One big advantage of the quark model in the present discussion is that it provides well defined predictions for all values of $w$, including the infinite 
mass limit where it can thus complement the lattice approach.
Among the many quark models, we are privileging here a particular approach
because of its remarkable properties, and especially those - but not only those -
which hold in the $m_Q=\infty$ limit. 

\subsection{General framework of Bakamjian-Thomas}\label{secBT}

The Bakamjian-Thomas approach may seem unfamiliar especially because the name
itself is unfamiliar, but nevertheless it has been used and developped and
promoted to a certain extent among nuclear physicists \cite{nuclear}
Its null-plane version has been extensively promoted in particle physics,
starting as it seems from M. Terent'ev \cite{terentev}.
 We believe that the form with
standard quantization on t=0 plane is nevertheless the closest to the
intuitions of the quark model. 
The main idea, starting from the researches of Foldy, is to formulate a
relativistic quantum mechanics with a fixed number of particle and an
instantaneous interaction. 
Of course, this is not fully possible, but already what is obtained is
significant and it seems to be the best one can do to implement relativity while
maintaining the three-dimensional spirit of the quark model.

Indeed, what is obtained is the full set of operators satisfying the Poincar\'e
algebra, which enables to define states in motion from states at rest; the latter
being described by wave functions which are eigenvectors of the mass operator.

One can define two sets of dynamical variables : one of global variables,
describing the state as whole, $\vec{P},\vec{R},\vec{S}$ ($\vec{R}=-i
\partial/\partial\vec{P}$ ), and one of internal variables, in fact the internal
momenta ${\vec{k}_i}$ (with $\sum_{i=1}^{n} \vec{k}_i = \bf{0} $), the conjugate positions, and the spins $\vec{s}_i$, with
global variables commuting with internal ones.

The mass operator M must be a rotation invariant function of internal variables
only, and then commutes with $\vec{P},\vec{R},\vec{S}$.

The Poincar\'e generators are constructed from the global operators  and the
interaction in a very simple form. i. e. they have the same form as for a free
particle, except that the mass (which appears in $H$ and the generator of
Lorentz transformations $\vec{K}$) is replaced by the mass operator M for the
bound states
\begin{eqnarray}
\vec{J}&=&\vec{P}\times -i\partial/\partial\vec{P} +\vec{S} \\
H&=&\sqrt{M^2+\vec{P}^2} \\
\vec{K}&=&1/2~[H,-i\partial/\partial\vec{P}]_{+} -\frac {\vec{P} \times \vec{S}}{H+M}
\end{eqnarray}

\subsection{Matrix elements of currrents}

One looses relativity when one writes matrix elements of some current
between multiquark states (see for details the paper \cite{covariant}). 

Multiquark states in motion are described by first writing a wave function in terms of
internal variables : 
\begin{eqnarray} \phi_{s_1,... s_n}(\vec{k}_1,...
\vec{k}_n) \qquad  
\end{eqnarray}
These are solutions of the eigenvalue equation $M \phi=m~\phi$. Then, the full wave
functions in terms of the usual one-particle variables 
\begin{eqnarray}
\Psi^{(\textbf{P})}_{s_1,... s_n}(\vec{p}_1,... \vec{p}_n) \qquad  \qquad \sum_{i=1}^{n} \vec{p}_i = \vec{P}
\end{eqnarray}
are obtained  
\begin{arabiclist}
\item by including a plane wave $\delta(\sum_i \vec{p}_i-\vec{P})$, 
\item by expressing the internal variables in terms of the one-particle variables,
which is obtained by a free-quark boost operation (with quark energy
$p_i^0=\sqrt{|\vec{p_i}|^2+m_i^2}$), including Wigner rotation of spins 
$s_1,... s_n$ and Jacobian factors ensuring unitarity. 
\end{arabiclist}
Whence the matrix element of a one-body operator.

If one chooses as current operators standard one quark
operators like $1_i,\sigma_i,...$ for simplicity and to comply with the basic
notion of additivity, then such matrix elements do not respect covariance. In
fact, only multibody current operators could implement the covariance. Neither
do the one-body currents satisfy the standard Ward identities like vector current
conservation.

The remarkable fact is that in the infinite mass limit of one quark, the main
difficulties disappear : one finds covariance, current conservation, and also
many properties required in this limit : Isgur-Wise scaling, normalisation
$\xi(w)=1$, a certain set of sum rules, namely those implying only the
Isgur-Wise functions (like the ones of Bjorken and Uraltsev)\footnote{On the
other hand, those implying also the energy levels (like the Voloshin sum rule),
are not exactly satisfied}. Also, one finds that in this limit, there is in fact
equivalence with a null plane formalism, with a suitable relation between the
wave functions at rest and on the null plane.

It is quite possible to calculate in the BT approach at physical masses.
However, for the present, in view of the above many difficulties concerning the
current matrix elements, and of the many advantages of the heavy mass limit, it
is logical to prefer using this limit, with the procedure described in 
section~\sref{historical} to connect it with phenomenology, keeping in mind that
non neglible ${\cal O}(1/m_Q)$ corrections are expected, especially for the
transition to $0^+$ (see subsection \sref{SL}).

\subsection{Choice of the potential model} 
To fix entirely the model, one needs of course to specify the mass operator $M$,
which means the potential model or wave equation at rest. i.e., we choose first
the general structure:
\begin{eqnarray}
M=\sum_i \sqrt{\vec{k}_i^2+m_i^2} + V
\end{eqnarray}
Then, we make specifically the choice of the model of Godfrey et Isgur  
(GI)~\cite{Godfrey:1985xj} for the
reason that it has been tested with success on a very large range of states,
which is less the case for others. We have checked that the
qualitative conclusions are general, by making calculations with other models.
We think however that in the present problem, where the wave function matters
much, the model of Godfrey et Isgur should be preferred also for the
following reason : we have noted a sensitivity to the short-distance behavior of
the potential for $\tau_{1/2}(w=1)$. With the potential of Veseli and Dunietz~\cite{veseli},
which has a strong Coulomb-like potential, one obtains a notably smaller value.
The GI model has a duely softened Coulomb-like potential.

In summary, we claim to use a relativistic approach to the quark model.
The meaning is twofold : 
1) one aspect is the relativistic boost, which allows to treat large velocities of each hadron as a whole.
2) one other is the choice of the internal kinetic energy. It means that the internal velocities may be large. 

Still other relativistic aspects, which we do not detail, have been  included in
the potential $V$ by Godfrey and Isgur. 

\section{Fundamental methods for QCD : lattice QCD}
There has not been many lattice caculations of $B \to D^{\ast\ast}$ decay
amplitudes. In the infinite mass limit only the calculation at $w=1$ seems
doable. As we shall see the results are in good agreement with the
prediction of the Bakamjian-Thomas model. 

Only recently the study at finite $b$ and $c$ mass has been undertaken. They
have to overcome very noisy signals and only preliminary results can now be 
presented. 

\subsection{Lattice calculation in the infinite mass limit} 
A preliminary calculation~\cite{Becirevic:2004ta} was performed to test a new
method. The infinitely massive quarks are described on the lattice by Wilson
lines in the time direction. The heavy quark is at rest. Thus one can only
consider the $w=1$ case in which the initial and final heavy quarks are both at
rest. However, the weak interaction matrix element vanishes at zero recoil. We 
overcome this problem using formulae from Leibovitch et al.~\cite{Leibovich:1997em} 
\begin{eqnarray}\label{scalaire}
\langle H^\ast_0(v) | A_i D_j  | H(v)\rangle  =  i \, g_{ij}\left (M_{H^{\ast}_0} -
 M_{H}\right)  \tau_{\frac 1 2}(1). 
 \end{eqnarray}
and 
\begin{eqnarray}\label{tenseur}
\langle H^\ast_2(v) | A_i D_j  | H(v)\rangle  = - i \sqrt{3}\left (M_{H^{\ast}_2} - 
M_{H}\right)
\tau_{\frac 3 2}(1) \epsilon^\ast_{ij}\;.
\end{eqnarray}
where the $H, H^\ast_0, H^\ast_2$ are the pseudoscalar, scalar and tensor mesons
in the infinite mass limit (the mass differences are finite) and 
$\epsilon^\ast_{ij}$ is the polarisation tensor of $H^\ast_2(v)$ and $v$ is the four-velocity. 
$A_i$ is the axial vector and $D_j$ is the covariant derivative which we know
how to discretize on the lattice.
Let us skip the details of the lattice calculation. 
The result of this ``quenched" lattice calculation~\cite{Becirevic:2004ta} was
\begin{eqnarray}\label{resultquenched}
\tau_{1/2}(1) = 0.38(4)(?)\quad{\rm and}\quad \tau_{3/2}(1) = 0.53(8)(?)\;,
\end{eqnarray}
where the question mark represent unknown systematic errors.
A similar calculation was performed later~\cite{Blossier:2009vy, Blossier:2009eg}
using gauge configurations of the European 
twisted mass collaboration with $N_f=2$ (two light quarks in the see). The
calculation has used several values of the light quark masses and one lattice
spacing. The result is
\begin{eqnarray} \label{final} 
\tau_{1/2}(1) &=& 
0.296(26) \quad , \quad \tau_{3/2}(1) =  0.526(23) \\ 
 \frac{\tau_{3/2}(1)}{\tau_{1/2}(1)} &=& 1.6 \ldots 1.8
\quad , \quad  \Big|\tau_{3/2}(1)\Big|^2 - \Big|\tau_{1/2}(1)\Big|^2 \approx
 0.17 \ldots 0.21 \end{eqnarray}
As already mentioned it is really close to the quark model result, 
\eref{eq:BTres}.
It is also noticeable that the ground state saturates about 80 \% of 
Uraltsev's sum rule, \eref{ur}.

\subsection{Lattice calculation with finite $b$ and $c$ masses}
The lattice calculation with finite masses is reported
in~\cite{Atoui:2013ksa,Atoui:2013sca}. The calculation is performed using also
the gauge configurations of the European twisted mass collaboration with
$N_f=2$. The calculation is performed with two lattice spacings, one value for
the charmed meson mass and three values for the ``$B$" meson : 2.5 GeV, 3. GeV
and 3.7 GeV. 
\subsubsection{The mass spectrum of $D^{\ast\ast}$ states}
The first issue is to check the mass spectrum predicted by lattice against
experiment and to try to relate them with the $j=1/2$ and $j=3/2$ states 
of the infinite mass limit. It is obvious that the $D^{\ast}_0$ ($D^{\ast}_2$)
corresponds the $j=1/2$ ($j=3/2$) in the ininifte mass limit. It is not so clear
for the $J=1$ states which mix  $j=1/2$ and $j=3/2$. This study has been
performed carefully by Kalinowski and Wagner~\cite{Wagner:2013laa}. They find
the first (second) excitation to be dominated by the $j=1/2$ ($j=3/2$)
contribution and they are identified to the broad $D_1(2430)$ (narrow $D_1(2420)$).
In~\cite{Atoui:2013ksa} an attemp to check the continuum limit for the
$D^{\ast}_0$  and $D^{\ast}_2$ masses is not successful.

\subsubsection{The $B \to D^\ast_0$ at zero recoil}
Using twisted quarks the $D^\ast_0$ can mix with the pseudocalar $D$. We need to
isolate both states using the generalised eigenvalue problem (GEVP). In 
section~\sref{SL} it is mentioned that the $B \to D^\ast_0$ at zero recoil, which
vanishes in the infinite mass limit, might give non vanishing results with
finite mass~\cite{Leibovich:1997em}. As discussed in~\sref{SL} the size
of this zero recoil contribution is obviously an important issue as it could
enhance significantly the $B \to D^\ast_0 l \nu$ branching ratio, improving the
agreement with experiment for the semileptonic decay but maybe spoiling the non
leptonic one\footnote{Notice that the non-leptonic decay $(\overline{B}^0 \to D_0^{1/2\
+}\pi^-)$ happens far from zero recoil, indeed at maximum recoil, but it is
difficult to believe that a strong zero recoil amplitude becomes very small at
maximum recoil.}, see~\sref{phenNL}. The preliminary study in~\cite{Atoui:2013ksa,Atoui:2013sca}
has thus concentrated on the zero recoil  $B \to D^\ast_0 l \nu$ decay, leaving
the full study for later.

\begin{table}[htb]
\tbl{Amplitude ratios $B\to D^\ast_0$ over $B\to D$ }
{\begin{tabular}{@{}ccccc@{}}\toprule
\hline
\bfseries{lattice spacing (fm)} & \bfseries{ratio $\boldsymbol{m_{b(1)}}$} & \bfseries{ratio $\boldsymbol{m_{b(2)}}$} & \bfseries{ratio $\boldsymbol{m_{b(3)}}$} & ratio at physical B\\
\colrule
0.085(3) & 0.33(5) & 0.24(4) & 0.16(4)&0.06(4) \\
\hline
0.069(2) & 0.40(4) &0.31(4) &0.20(4)&0.09(4)\\
\hline
continuum &0.55(16) & 0.45(17)& 0.29(18) &0.15(20)\\
\botrule
\end{tabular}}
\begin{tabnote} We give the amplitude ratios $B\to D^\ast_0$ over $B\to D$
 at zero recoil, averaged over what seems a good plateau : 6-9 (5-11) for 
 lattice spacing $a=0.085$ fm
($a=0.069$ fm). 
The $b$ quarks range from the lightest to heaviest 
from left to right : $m_B \simeq 2.5,3,3.7$ GeV. The right column 
corresponds to the extrapolation at the physical $B$ mass, 5.2 GeV.
The last line corresponds to the extrapolation to the continuum.  
\end{tabnote}\label{tab:ratioDscalD}
\end{table}
The results are
shown in~\tref{tab:ratioDscalD}. From these number we get several conclusions
\begin{itemlist} 
\item For the ``$B$ meson" masses used and finite lattice spacing the ratios
 are significant.
\item But they decrease with increasing $B$ mass leading to a small signal/noise 
ratio at the physical $B$ mass
\item The extrapolation to the continuum increases the noise leading to 
more than 100~\% uncertainty. 
\end{itemlist}
We can thus conclude that there is really a significant zero recoil 
 $B\to D^\ast_0$ over $B\to D$ amplitude ratio which however, when extrapolated
to the physical situation is too uncertain. The situation should improve using
a third smaller lattice spacing and larger statistics.
\subsubsection{Possible interpretation of $D^\ast_0$ as a scattering state :}

Mohler et al.~\cite{Mohler:2012NA} consider the possible interpretation
of the $D^\ast_0$ as a $D\,\pi$ scattering state. Indeed, what is 
identified as $D^\ast_0$ is a broad structure which is certainly strongly 
coupled to the $D\,\pi$ channel. Under this hypothesis, the study of the 
$B \to D^\ast_0 l \nu$ should be totally reconsidered.  
\begin{figure}[ht]
\centerline{\psfig{file=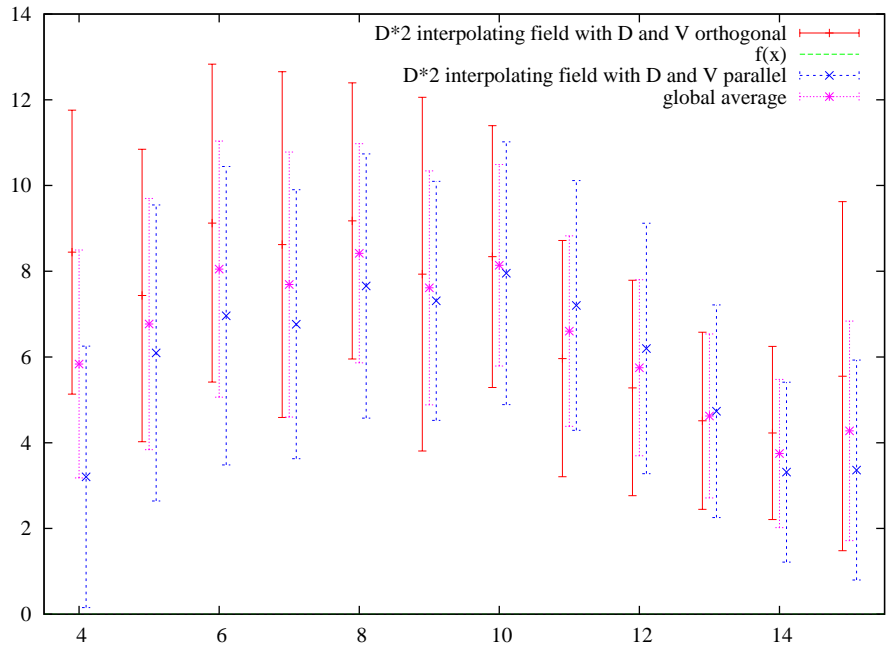}}
\caption{One example of ratio of the  $B \to D^*_2$ amplitude over the
value derived from the infinite mass limit, once the 
zero recoil has been subtracted. The ``B meson mass" is 2.5 GeV and the lattice
spacing 0.069 fm. The momentum of the B meson correponds to $w=1.3$.
The blue cross corresponds to the discrete
representation $E^+$ for the $ D^*_2$ interpolating field, $D_iV_i$, while the
red bar corresponds to the 
discrete representation $T_2^+$ i.e. $D_iV_j, j \ne i$. The global average is the
purple star.}
\label{fig:ratioB2D2}
\end{figure}
\subsubsection{The $B \to D^\ast_2$ amplitude compared to the infinite mass limit}
The $B \to D^\ast_2$ decay amplitude vanishes at zero recoil. This is obvious 
in the continuum since a spin 0 particle cannot decay into a spin 2 one via
Axial matrix elements which carry spin 0 or 1. However, when using twisted mass 
quarks, the parity symmetry is violated. The twisted quarks are gathered into
isospin doublets and even the heavy quarks are represented in fake ``isospins".
There exists an exact  symmetry valid at finite lattice spacing consisting in a
parity symmetry combined  with the flip of the isospin. Thus combining both
isospins one gets decay amplitudes which vanish at zero recoil. Of course
statistical fluctuations make them not zero. However these fluctuations at zero
recoil and non zero recoil are correlated. Thus we can reduce the fluctuations
by subtracting the zero recoil result from the non zero recoil result.  

As a benchmark we use the prediction of the infinite mass limit formula. We show
in \fref{fig:ratioB2D2} one example of such a ratio. A spin 2 meson has 5
possible polarisation. On a lattice these are decomposed into two different
discrete symmetry groups : $E^+$ and $T_2^+$. $E^+$ corresponds to appropriate
combinations of terms of the type $D_i\,V_i$ where $D_i$ is the discretised
covariant derivative and $V_i$ the vector current, and $T_2^+$ corresponds to an
apropriate combination of terms of the type $D_i\,V_j; j\ne i$,
see \cite{Atoui:2013ksa} for more details. 

\Fref{fig:ratioB2D2} shows that there is definitely a signal and the
agreement between $E^+$ and $T_2^+$ is rewarding. However the ratio is very
large. Since the infinite mass formula agrees rather well with experiment we
would expect a ratio closer to 1. This is not the case, the ratio is one order
of magnitude larger than 1 and this does not seem to improve when going towards 
the physical B mass neither towards continuum. The reason for this problem is
not yet understood.

\section{Experiment}

\subsection{Non leptonic versus semileptonic}

These respective processes present advantages and drawbacks with respect to each
other, so that, on the whole, they should be considered complementary. Indeed:

$\alpha$) On the one hand, semileptonic data have have a series of theoretical
advantages. They are directly connected with the matrix elements of the
currents, which are the true theoretical objects under study. Non leptonic
processes can be related to them only trough the additional assumption of
factorisation, which has not a very strong theoretical foundation, although it
seems to work well in similar decays like $B \to D^{(*)} \pi$. Moreover, in
Class III they are complicated by the additional diagram with $D^{**}$ emission.
And in Class I, $B \to D \pi \pi$ contains a non exotic, strongly resonating 
$\pi^+ \pi^-$ channel which must be extracted off.

$\beta$) One could also believe that the larger branching ratios of semileptonic
decays ($10^{-3}$ versus $10^{-4}$) should give them a big superiority in
statistics. However, this is overcompensated by far by the fact that the
detection efficiency of semileptonic decay is much worse. One has four bodies in
the final state instead of three, and a neutrino among them. Finally, the non
leptonic data are much more accurate. It is only from them that one has been
able to determine the mass and width of the broad states.

Note that one cannot consider the two processes as truly independent, because
of their relation through factorisation.  
\subsection{Narrow versus broad states}
 The narrow states $j=3/2$  have always shown quite a good consistency among
experimental measurements. And as already said, the theoretical
numbers have complied with experiment.

On the other hand, for the broad $j=1/2$, the experimental situation, instead of
becoming clearer, has fallen into a real state of confusion. Indeed, aside
from the attempts by Delphi at LEP, who found a very large contribution of the
broad $1^+$ in semileptonic decay, the higher luminosity B factories could be
expected to clear up the question; in fact, contrasting results have been
obtained by  Belle and Babar, with different situations for $0^+$ versus $1^+$
and for non leptonic versus semileptonic processes.

The fact that narrow states results are quite consistent, while those for broad
states are not, suggests that it is the broadness which could be responsible for
the difficulties encountered for these states. Indeed, they are very broad :
around $300$ MeV. It is also why we shall present separately the experimental
result for
the two type of states.
\subsection{Results}

We begin, in an unusual way, with the non leptonic decay because the situation
is clearer, and also to insist on the often underestimated necessity to take
them into account for the discussion of the $j=1/2$ case. Since there is
compatibility between Belle and Babar where both have measured, we quote only
in \tref{tab:dsstarpi} their averaged results. 

\begin{table}[!htb]
\tbl{Nonleptonic data $j=3/2, 1/2$}
{\begin{tabular}{@{}ccc@{}}\toprule
\hline
 Decay channel  & $B_d^{0},~Class~I$  & $B^{+},~Class~III $ \\
\hline
 $\overline{D}_2^* \pi^+$& $(0.49 \pm 0.07)\times 10^{-3}$& $(0.82 \pm 0.11)\times 10^{-3}$ \\
\hline
 $\overline{D}_1^{3/2} \pi^+$& $(0.82_{-0.17}^{+0.25})\times 10^{-3} $& $(1.51 \pm 0.34)\times 10^{-3}$ \\
\hline 
 $\overline{D}_1^{1/2} \pi^+$& $<1 \times 10^{-4} $& $(0.75 \pm 0.17)\times 10^{-3}$ \\
\hline
 $\overline{D}_0^* \pi^+$& $(1.0 \pm 0.5)\times 10^{-4}$& $(0.96 \pm 0.27)\times 10^{-3}$ \\
\hline
\botrule
\end{tabular}}
\begin{tabnote}Non leptonic branching ratios $B \to D^{\ast\ast} \pi$ averaged
between Belle and Babar~\cite{ref:babar_dpipich,ref:belle_dpipich,ref:babar_dpipine,
ref:belle_dpipine1,ref:belle_dpipine2}.The final strong decay branching fractions have been 
taken as detailed footnote 6 of~\cite{proposal}.  
\end{tabnote}\label{tab:dsstarpi}
\end{table}

One notes the remarkable fact that in Class I, $j=1/2$  are one order of
magnitude smaller than $j=3/2$, while all have the same order of magnitude
$10^{-3}$ in Class III. This is explained clearly by theory, see above,
\sref{phenNL}. 

For the Class I, the following warnings must be made :
\begin{itemize}

\item the large uncertainty ($50\%$) in the $0^+$ case, is due to the fact that Babar
has an appreciably larger value, but with a very large uncertainty. This very
large uncertainty is itself due to the uncertainty in the extraction of a non
resonant contribution. 

\item the $1^+, j=1/2,3/2$ ($D^* \pi$ channel) have been measured only by Belle.
\end{itemize}
\begin{table}[!htb]
\tbl{Semileptonic data $j=3/2$}
{\begin{tabular}{@{}ccc@{}}\toprule
\hline
 Decay channel  & $Belle$  & $Babar$ \\
\hline
 $\overline{D}_2^* ~l \nu$ & $(0.54 \pm 0.12)\times 10^{-2}$& $(0.39 \pm 0.11)\times 10^{-2}$ \\
\hline
 $\overline{D}_1^{3/2}~ l \nu$ & $(0.93 \pm 0.22)\times 10^{-2} $& $(0.64 \pm 0.15)\times 10^{-2}$ \\
\hline \botrule
 \end{tabular}}
\begin{tabnote}
Semileptonic branching ratios for B semileptonic decay into narrow states, from
Belle and Babar. See HFAG~\cite{ref:hfag} and references therein. The final strong
 decay branching fractions have been taken as detailed footnote 6 of~\cite{proposal}.
\end{tabnote}\label{tab:SLnarrow}
\end{table}
For semileptonic decays we quote first the results of $j=3/2$ (narrow states) in
\tref{tab:SLnarrow} : Obviously, the two experiments are here quite
compatible. 

\begin{table}[!htb]
\tbl{Semileptonic data $j=1/2$}
{\begin{tabular}{@{}ccc@{}}\toprule
\hline
 Decay channel  & $Belle$  & $Babar$ \\
\hline 
 $\overline{D}_1^{1/2}~ l \nu$& $<1.05 \times 10^{-3} $ & $(0.40 \pm 0.05)\times 10^{-2}$ \\
\hline
 $\overline{D}_0^* ~ l \nu$& $(0.36 \pm 0.09)\times 10^{-2}$& $(0.42 \pm 0.09)\times 10^{-2}$ \\
\hline\botrule
\end{tabular}}
\begin{tabnote}
Semileptonic branching ratios for B semileptonic decay into broad states, from
Belle and Babar. See HFAG~\cite{ref:hfag} and references therein. The final strong
 decay branching fractions have been taken as detailed footnote 6 of~\cite{proposal}. 
\end{tabnote}\label{tab:SLbroad}
\end{table}
The semileptonic decay into $j=1/2$ from Babar and Belle is reported in 
\tref{tab:SLbroad}~: There is a strong disagreement
for the $1^+(1/2)$ between the two experiments while the results for $0^+$ are 
quite compatible

About $0^+$ : one could say that in view of the agreement between the two
experiments in the semileptonic case, we might trust this large result. But
given the rather direct relation between semileptonic and Class I non leptonic
decays, and the strong suppression of $0^+$ in non leptonic (see the above
\tref{tab:dsstarpi}), this conclusion would be very doubtful.

The general conclusion could be that, not surprisingly, the very broad states
raise more difficulties precisely because of their broadness.

\section{Conclusion : discussion on theory and experiment, prospects}

\subsection{Discussion on theory and experiment}
One must say that the non leptonic data, \tref{tab:dsstarpi}, seem to
support strongly theoretical expectations coming from the $m_Q
\to \infty$ approximation:

1) suppression of $j=1/2$ with respect to $j=3/2$ in Class I, by one order of
magnitude. The agreement with~\eref{GIclassI} is semiquantitative for both $j$,
and good if summing over the members of the $j$ multiplets. 

2) This suppression is no more present in Class III, due to the diagram with
$D^{**}$ which is large in the $j=1/2$ case, but not for $j=3/2$, also from a
statement of $m_Q \to \infty$ limit. Indeed~\cite{jugeau} one observes that the
$j=1/2$ are strongly enhanced in class III, by one order of magnitude, and the
$j=3/2$ much more slightly (the two diagrams can be shown to add
constructively).

The $j=3/2$ semileptonic data are much larger, as expected from the $m_Q \to \infty$
approximation, and in quantitative agreeement when averaging over the members of the multiplet.

The only definite discrepancy with theory in the $m_Q \to \infty$ limit is in the
semileptonic case, for the $j=1/2, 0^+$ state, where the two experiments agree. It could be understood as coming
from the
broadness of the states and the lower detection efficiency of semileptonic measurements, rendering the identification of the resonance still more
difficult (much less observed events) . 

Of course, part of the discrepancy could be due to the $1/m_Q$ corrections being large. This can be 
suggested  by the finding of Leibovich et al., that precisely this transition
can suffer more $1/m_Q$ enhancement. This effect could combined with the fact
that lattice QCD finds a somewhat larger $\tau_{1/2}(1)$ than our quark model
(see \sref{well}). But then, it would remain to explain the non
leptonic data : why then $0^+$ seems so small in Class I. Therefore, still something would remain problematic on one or another side of experiment. 

Another problem is the large Babar result for the broad $1^+,1/2$  (contrasting with the small one from Belle). It would be a real worry if confirmed, because in this case one does not expect any serious enhancement  according to the calculations of Leibovich et al. (the conclusion is the same in the quark model \cite{finite}).

\subsection{Prospects}

\begin{itemize}

\item The most urgent problem seems the experimental one :

1) to solve the discrepancy between the two B factory, see \tref{tab:SLbroad},
concerning the semileptonic decay to $1^+(j=1/2)$
 
2) to clarify the discrepancy which seems to exist between the semileptonic and
the Class I non leptonic data for the $0^+$ ($j=1/2$) (\tref{tab:dsstarpi}
and \tref{tab:SLbroad}), if we believe
factorisation. There is a strong suppression in the latter with respect to the
$j=3/2$, while  $0^+$ is of the same order as $j=3/2$ in semileptonic. It would
be already very useful to reduce the uncertainty of Babar measurement of the
Class I non leptonic decay for the $0^+$, and to have the Babar measurement of
the $1^+(j=1/2)$ partner. 

It would also be very useful to complement the study of the broad states by the
one of their strange counterparts, which are narrow, and therefore much 
easier to identify. This is proposed in the article~\cite{proposal}.

On the other hand, on the theoretical side :

\item since one can ``fear" unexpectedly large  $1/m_Q$ corrections in the $0^+$,
lattice QCD at finite mass should be developped as much as possible. For the
moment lattice calculations at finite mass make it very likely that a
significant zero recoil contribution to $B \to D^\ast_0$ is there. However 
the extrapolation to the physical situation leads to more than 100 \%
uncertainty, \tref{tab:ratioDscalD}. This of course is not the case for
$B \to D^\ast_2$ but there the ratio of the finite mass signal over the infinite
mass estimate is larger than expected and than experiment for an unknown 
reason. The progress will come from using additionally a smaller lattice spacing
to make safer the continuum limit, to look in detail into the momentum
dependance of all these decays and to chase possible remaining artefacts.

\item  since one needs anyway other approaches, mainly the quark model, to
understand things at large $w$ (for example $q^2=0$ for the pionic weak
transition), one should try to improve estimates of  $1/m_Q$ corrections.

\item Finally, to justify the longstanding efforts dispensed on the problem,
both from the part of theorists and of experimentalists, it must be said why it
seems so important to clarify the situation for $L=1,~j=1/2$. It is because, on
the whole, a serious issue is the {\bf phenomenological relevance of the 
$m_Q=\infty$ approach in $b \to c$ decay}. This limit has been considered very
attractive because allowing several important new theoretical statements, in
particular the sum rules, which deeply involve theses states ; and at the same
time, it has met quite encouraging phenomenological successes in $B \to D^{(*)}$
and $B \to D^{**}, L=1,~j=3/2$ transitions, as well as in the strong decays $D^{**} \to D^{(*)} \pi$ with both $j=1/2$ and $j=3/2$.

Of course it is expected that the results of $m_Q=\infty$ are a better
approximation for the $j=3/2,2+$ state than for the other cases. Indeed, in the latter cases, the no recoil amplitude at $m_Q=\infty$ vanishes contrary to general expectation, and a non zero value should be present, resulting from $1/m_Q$ effects. Indeed, theoretical expectations from certain identities lead to possible large effects at order ${\cal O}(1/m_Q)$ in the $1+,3/2$ and $0^+$ cases. Unluckily,
lattice are not yet able to answer clearly to their size at physical masses, and whether they could fill the large gap with presently observed value for the $0^+$ in the semileptonic decay . Moreover, such a large non zero value at no  recoil
which would mean an S-wave between the final $D^{\ast\ast}$ and the $l \nu$ system, is to be confronted
to the success of the $m_Q=\infty$ approximation for the class
I non-leptonic decays. Can this S-wave effect be damped in the maximal recoil
kinematics which is that of the  $D^{\ast\ast} \pi$ ? Models rather point to a
soft variation of the corresponding effect.
\item
This longstanding efforts are also finally justified by the important side 
effect that the  decays considered here have on the estimate of $V_{cb}$.
\end{itemize}


\begin{thebibliography}{99}

\bibitem{memorino1}
Memorino on the `$1/2$ versus $3/2$ puzzle' in $\bar B \to l~\bar \nu X(c)$.
I.I. Bigi, B. Blossier, A. Le Yaouanc, L. Oliver, O. Pene, J.-C. Raynal, A. Oyanguren, P. Roudeau.
e-Print: hep-ph/0512270

\bibitem{memorino2}
I.I. Bigi, B. Blossier, A. Le Yaouanc, L. Oliver, O. Pene, J.-C. Raynal, A. Oyanguren, P. Roudeau.
Memorino on the `$1/2$ versus $3/2$ puzzle' in $\bar B \to l~\bar \nu X(c)$: A Year Later and a Bit Wiser.
Published in Eur.Phys.J.C52:975-985,2007.
e-Print: arXiv:0708.1621 [hep-ph]

\bibitem{proposal} 	
Proposal to study $Bs\to D_{sJ}$ transitions
Damir Becirevic, Alain Le Yaouanc, Luis Oliver, Jean-Claude Raynal (Orsay, LPT), Patrick Roudeau (Orsay, LAL), Justine Serrano (Marseille, CPPM). Jun 2012. 13 pp.
Published in Phys.Rev. D87 (2013) 5, 054007
e-Print: arXiv:1206.5869 [hep-ph] 

\bibitem{Isgur:1990jf}
  N.~Isgur and M.~B.~Wise,
  Phys.\ Rev.\ D {\bf 43} (1991) 819.

\bibitem{duality} Exact duality and Bjorken sum rule in heavy quark models a la Bakamjian-Thomas,
A. Le Yaouanc, L. Oliver, O. Pene, J.C. Raynal
    Phys.Lett. B386 (1996) 304-314
        e-Print: hep-ph/9603287 

\bibitem{Uraltsev:2000ce}
  N.~Uraltsev,
  Phys.\ Lett.\ B {\bf 501} (2001) 86
  [hep-ph/0011124].

\bibitem{LeYaouanc:2001vu}
  A.~Le Yaouanc, L.~Oliver, O.~Pene, J.~C.~Raynal and V.~Morenas,
  PoS HEP {\bf 2001} (2001) 082
  [hep-ph/0110372].

\bibitem{uraltsev-nous} Uraltsev sum rule in Bakamjian-Thomas quark models,
  A. Le Yaouanc, L. Oliver, O. Pene, J.C. Raynal , V. Morenas, 
    Phys.Lett. B520 (2001) 25-32
    e-Print: hep-ph/0105247 
 
  
\bibitem{Morenas:1997nk}
  V.~Morenas, A.~Le Yaouanc, L.~Oliver, O.~Pene and J.~C.~Raynal,
  Phys.\ Rev.\ D {\bf 56} (1997) 5668
  [hep-ph/9706265].
  
\bibitem{Godfrey:1985xj}
  S.~Godfrey and N.~Isgur,
  Phys.\ Rev.\ D {\bf 32} (1985) 189.

\bibitem{ref:pdg}
K. Nakamura et al. (Particle Data Group), J. Phys. G 37, 075021 (2010) 
and 2011 partial update for the 2012 edition.

\bibitem{delphi}
  J.~Abdallah {\it et al.}  [DELPHI Collaboration],
  Eur.\ Phys.\ J.\ C {\bf 45} (2006) 35
  [hep-ex/0510024].


\bibitem{Leibovich:1997em}
  A.~K.~Leibovich, Z.~Ligeti, I.~W.~Stewart and M.~B.~Wise,
  Phys.\ Rev.\ D {\bf 57} (1998) 308
  [hep-ph/9705467].
 
\bibitem{finite} Finite mass corrections to $B \to D^{(*)}, D^{**}$ transitions
in the Bakamjian-Thomas approach,A.~Le Yaouanc, L.~Oliver, J.~C.~Raynal (to be
published).
 
 \bibitem{neubert} Theoretical analysis of $\bar B \to D^{\ast\ast} \pi$ decays.
M. Neubert.
Published in Phys.Lett.B418:173-180,1998.
e-Print: hep-ph/9709327
 
\bibitem{jugeau} 
The Decays $\bar B \to D^{\ast\ast} \pi$ and the Isgur-Wise functions
$\tau(1/2)(w)$, $\tau(3/2)(w)$. F. Jugeau, A. Le Yaouanc, L. Oliver, J.-C.
Raynal, Published in Phys.Rev.D72:094010,2005. e-Print: hep-ph/0504206

\bibitem{nuclear} B. Keister, W. Polyzou, Adv. Nucl. Phys.{\bf 20}, 225 (1991).

\bibitem{terentev} M. Terent'ev Sov. J.Nucl. Phys. {\bf 24}, 106 (1971).

\bibitem{covariant} Covariant quark model of form-factors in the heavy mass limit,
A. Le Yaouanc, L. Oliver, O. Pene, J.C. Raynal, 
    Phys.Lett. B365 (1996) 319-326
    
\bibitem{veseli}  S. Veseli and I.Dunietz  Phys. Rev. D54 6803 (1996)  

\bibitem{Becirevic:2004ta}
  D.~Becirevic, B.~Blossier, P.~Boucaud, G.~Herdoiza, J.~P.~Leroy, A.~Le Yaouanc, V.~Morenas and O.~Pene,
  Phys.\ Lett.\ B {\bf 609} (2005) 298
  [hep-lat/0406031].
  
\bibitem{Blossier:2009vy}
  B.~Blossier {\it et al.}  [European Twisted Mass Collaboration],
  JHEP {\bf 0906} (2009) 022
  [arXiv:0903.2298 [hep-lat]].
  
\bibitem{Blossier:2009eg}
  B.~Blossier {\it et al.}  [ETM Collaboration],
  PoS LAT {\bf 2009} (2009) 253
  [arXiv:0909.0858 [hep-lat]].
  
\bibitem{Atoui:2013ksa}
  M.~Atoui, B.~Blossier, V.~Morenas, O.~Pene and K.~Petrov,
  arXiv:1312.2914 [hep-lat].
  
\bibitem{Atoui:2013sca}
  M.~Atoui,
  arXiv:1305.0462 [hep-lat].
  
\bibitem{Wagner:2013laa}
  M.~Wagner and M.~Kalinowski,
  arXiv:1310.5513 [hep-lat].
  
  
 \bibitem{Mohler:2012NA}
D.~Mohler, S.~Prelovsek and R.~M.~Woloshyn,
Phys.\ Rev.\ D {\bf 87}, no. 3, 034501 (2013).
[arXiv:1208.4059 [hep-lat]].

\bibitem{ref:babar_dpipich}
B. Aubert {\it et al.}, BaBar Collaboration, Phys. Rev. {\bf D79} 112004, 2009.
  [arXiv:0901.1291 [hep-ex]].


\bibitem{ref:babar_dpipine} 
P. del Amo Sanchez  {\it et al.}, BaBar Collaboration, 
SLAC-PUB-14203, e-Print: arXiv:1007.4464 [hep-ex].
  PoS ICHEP {\bf 2010} (2010) 250

\bibitem{ref:belle_dpipich}
K. Abe {\it et al.}, Belle Collaboration, Phys. Rev. Lett. {\bf 94} 221805, 2005.
 K.~Abe {\it et al.}  [Belle Collaboration],
  [hep-ex/0410091].

\bibitem{ref:belle_dpipine1}
K. Abe {\it et al.}, Belle Collaboration, Phys. Rev. {\bf D69} 112002, 2004.
  [hep-ex/0307021].

\bibitem{ref:belle_dpipine2}
A.~Kuzmin  {\it et al.}, Belle Collaboration, Phys. Rev. {\bf D76} 012006, 2007.
 {\it et al.}  [Belle Collaboration],
  [hep-ex/0611054].

\bibitem{ref:hfag}
  D.~Asner {\it et al.}  [Heavy Flavor Averaging Group Collaboration],
  arXiv:1010.1589 [hep-ex].

\end{thebibliography}
\end{document}